\documentclass{elsart}
\usepackage{amssymb}
\usepackage{graphicx}

\begin{document}
\begin{frontmatter}


\title{Improved limit on electron neutrino charge radius through a new
  evaluation of the weak mixing angle}

\author[cinvestav]{J. Barranco},
\ead{jbarranc@fis.cinvestav.mx}
\author[cinvestav]{O. G. Miranda},
\ead{Omar.Miranda@fis.cinvestav.mx}
and
\author[mpi,izmiran]{T. I. Rashba\corauthref{cor}}
\corauth[cor]{Corresponding author.}
\ead{timur@mppmu.mpg.de}

\address[cinvestav]{Departamento de F\'{\i}sica, Centro de
  Investigaci{\'o}n y de Estudios Avanzados del IPN, Apdo. Postal
  14-740 07000 M\'exico, D F, M\'exico}

\address[mpi]{Max-Planck-Institut f\"ur
Physik (Werner-Heisenberg-Institut), F\"ohringer Ring 6, 80805
M\"unchen, Germany}
\address[izmiran]{IZMIRAN,
142190, Troitsk, Moscow region, Russia}

\begin{abstract}
  We have obtained a new limit on the electron neutrino effective
  charge radius from a new evaluation of the weak mixing angle by a
  combined fit of all electron-(anti)neutrino electron elastic
  scattering measurements. Weak mixing angle is found to be
  $\sin^2\theta_W=0.259\pm0.025$ in the low energy regime below 100
  MeV.  The electron neutrino charge radius squared is bounded to be
  in the range $-0.13\times10^{-32}$~cm$^{2} < \langle r_{\nu_e}^2
  \rangle < 3.32\times10^{-32}$~cm$^{2}$ at 90\% C.L.  Both results
  improve previously published analyses. We also discuss perspectives 
  of future experiments to improve these constraints.
\end{abstract}

\end{frontmatter}

\section{Introduction}

The search for neutrino electromagnetic properties is as old as
neutrino theories~\cite{Pauli:1930pc}. The electromagnetic interaction
of Dirac neutrino is described by four form-factors: vector, axial
vector, magnetic and electric
ones~\cite{Nieves:1981zt,Shrock:1982sc}. In the low energy limit the
first two form-factors are related with neutrino effective vector and
axial vector (anapole) charge radii, the magnetic and electric form
factors are related with neutrino magnetic and dipole moments,
respectively.  In case of Majorana neutrinos only axial vector form
factor and off-diagonal magnetic and electric dipole moments are
non-zero, but still one cannot distinguish Dirac from Majorana
neutrinos in these interactions because of the relativistic nature of
neutrinos.

The gauge-invariant definition of the neutrino effective charge radius
as a physical observable has been discussed since a long
time~\cite{Bardeen:1972vi,Lee:1973fw,Sarantakos:1982bp,Degrassi:1989ip,Fujikawa:2003ww,Fujikawa:2003tz}.
Recently it was shown that gauge-dependent terms cancel each other and
therefore neutrino charge radius can be defined as a gauge-invariant
physical
observable~\cite{Bernabeu:2000hf,Bernabeu:2002pd,Bernabeu:2002nw,Bernabeu:2003xj};
however, theoretical expectations~\cite{Bernabeu:2000hf} for electron
neutrino charge radius are one order of magnitude smaller than present
experimental limits. On the other hand it was shown in
Ref.~\cite{Hirsch:2002uv} that it is not possible to reach a good
sensitivity to neutrino charge radius through astrophysical and
cosmological observations and therefore it can be constrained only in
terrestrial experiments.

At the low energy scale, besides atomic parity
violation~\cite{Bennett:1999pd} and Moller
scattering~\cite{Anthony:2005pm}, neutrino-electron scattering
experiments are sensitive to the weak mixing angle and they have also
been used to search for neutrino effective charge radius. Searches for
a non-zero neutrino magnetic moments have also been a challenge in
this kind of
experiments~\cite{Daraktchieva:2005kn,Wong:2006nx,Beda:2007hf}.

Recently, the need for a precise measurement of the weak mixing 
angle at low energies has been encouraged by the NuTeV
collaboration~\cite{Zeller:2001hh}, in which a $3\sigma$ deviation of
the Weinberg angle from the Standard Model prediction was found in
deep inelastic neutrino scattering. 

In this paper we introduce a new limit on the  weak mixing angle,
$\sin^2\theta_W$, and on the electron neutrino effective charge radius,
$\langle r_{\nu_e}^2 \rangle$, obtained from the combined analysis of all
available low energy (anti)neutrino-electron elastic scattering 
experiments.

We have analyzed all available measurements of the
(anti)neutrino-electron scattering from the following reactor and
accelerator experiments: first measurement of neutrino-electron
scattering made by Reines, Gurr and Sobel
(Irvine)~\cite{Reines:1976pv}, the Kurchatov institute group at the
Krasnoyarsk reactor~\cite{Vidyakin:1992nf}, the group from Gatchina at
the Rovno reactor~\cite{Derbin:1993wy}, MUNU at the Bugey
reactor~\cite{Daraktchieva:2005kn}, LAMPF~\cite{Allen:1992qe} and
LSND~\cite{Auerbach:2001wg}.

We haven't included into our analysis the recent reactor neutrino results 
from
TEXONO~\cite{Wong:2006nx} and GEMMA~\cite{Beda:2007hf} experiments, which 
have
put very strong limits on neutrino magnetic moments, $\mu_\nu < 7.4 \times
10^{-11}\mu_B$ and $\mu_\nu < 5.8\times 10^{-11}\mu_B$ at 90\% CL,
respectively.
Both experiments are working at very low energy thresholds searching
for neutrino magnetic moments and are not yet sensitive to the weak
mixing angle and neutrino charge radius. 
The discussion of theoretical and experimental issues on muon- and
tau-neutrino effective charge
radii~\cite{Bernabeu:2000hf,Hirsch:2002uv} is also out of
the scope of this paper.

The paper is organized as follows: in Section II we derive limits to
the weak mixing angle and to the neutrino charge radius, in Section
III we discuss future perspectives of precise measurements at reactor
neutrino experiments. The Summary is presented in Section
IV.

\section{Limits on weak mixing angle and neutrino charge radius}

The differential weak cross section for electron (anti)neutrino scattering off
electron is given by
\begin{equation}
  \frac{d\sigma}{dT}=\frac{G_F^2 m_e}{2\pi}
  \left[\left(g_V+g_A\right)^2+\left(g_V-g_A\right)^2\left(1-\frac{T}{E_\nu}
    \right)^2 
-\left(g_V^2-g_A^2\right)\frac{m_e T}{E_\nu^2}\right]\,,
\end{equation}
here $E_\nu$ is the incoming neutrino energy, $T$ is the recoil
electron energy.  The vector and axial weak couplings are given as
$g_V=1/2+2\sin^2\theta_W$ and $g_A=1/2$ for neutrinos, while in the
case of anti-neutrinos $g_A=-1/2$. 

\begin{figure}
\includegraphics*[width=\columnwidth]{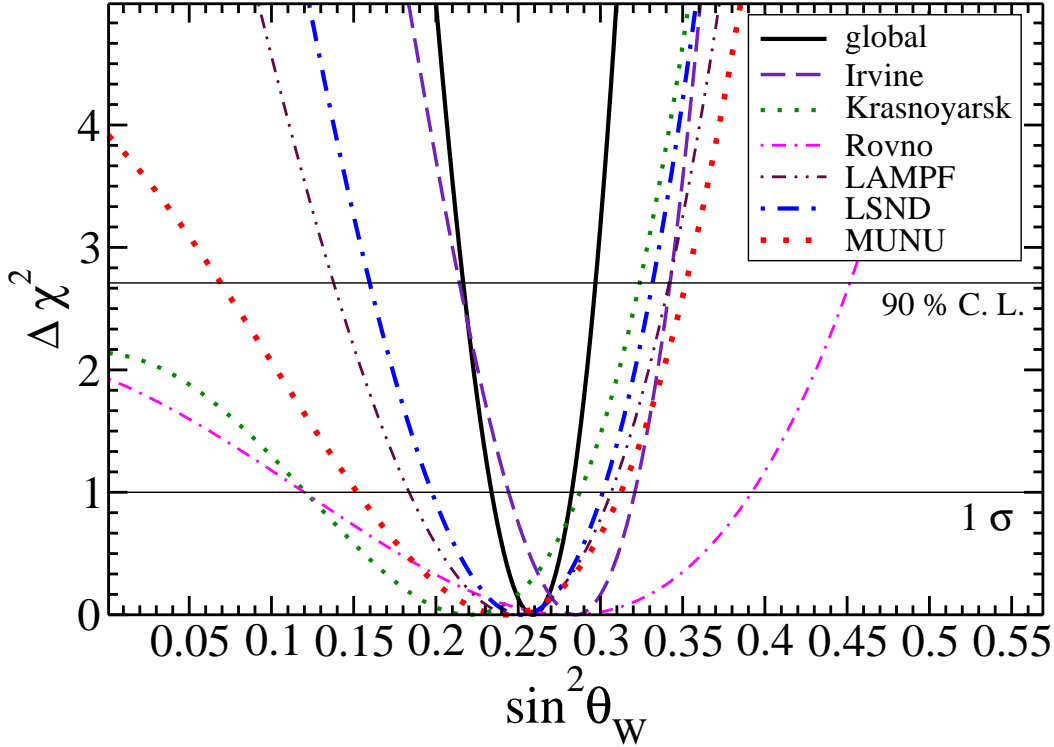}
\caption{Global $\Delta \chi^2$ for weak mixing angle
obtained from all discussed $\nu_e e$ and $\bar{\nu}_e e$
  scattering experiments are plotted.
The contribution of each experiment to the $\Delta
  \chi^2$ is also shown.
}\label{fig:sin}
\end{figure}

We will concentrate first in the determination of the weak mixing
angle from (anti)neutrino electron scattering experiments and we will
discuss the neutrino charge radius in detail afterwards.  To extract
the allowed region for the weak mixing angle from the experimental
data we have done a $\chi^2$ analysis (details can be found in
Ref.~\cite{Barranco:2005ps}) for each experiment and then we have
performed a global fit. In particular, we have consider two different
$\nu_e e$ experiments, LAMPF and LSND. In both cases we have
confronted the theoretical cross section with the reported
measurement to obtain our value of $\sin^2\theta_W$. We also considered the
experimental results reported in four different reactor experiments.
In this case we compute the total antineutrino cross section
\begin{equation}
\sigma =  \int dT' \int dT \int dE_\nu 
                    \frac{d\sigma}{dT} \lambda (E_\nu) R(T,T')
\label{diff:cross:sec-sm}
\end{equation}
with  $\lambda (E_\nu)$ the neutrino energy spectra and $R(T,T')$ 
the detector energy resolution function. 
We use an anti-neutrino 
energy spectrum given by 
\begin{equation}
\lambda(E_{\nu}) = \sum_{k=1}^4 a_k \lambda_k(E_\nu) ,
\label{diff:cross:sec-NSI}
\end{equation}
where $a_k$ is the abundance of $^{235}$U ($k=1$), $^{239}$Pu
($k=2$), $^{241}$Pu ($k=3$) and $^{238}$U ($k=4$) in the
reactor, $\lambda_k(E_\nu)$ is the corresponding neutrino energy
spectrum which we take from the parametrization given
in~\cite{Huber:2004xh}, with the appropriate fuel composition. For
energies below $2$~MeV there are only theoretical calculations for the
antineutrino spectrum which we take from Ref.~\cite{Kopeikin:1997ve}.
For the case of the Irvine experiment (that reports two energy bins)
we prefer to use the neutrino energy spectrum used by the
experimentalists at that time~\cite{Avignone}. With this cross section
evaluation we can perform a $\chi^2$ analysis either comparing with the
expected number of events, for the case of the MUNU collaboration, or
with the reported cross section in the rest of the experimental results.
We neglect correlations between experiments; this is a good
approximation as the only possible correlation comes from the reactor
neutrino energy spectrum, estimated to be less than
2\%~\cite{Huber:2004xh}, small in view of the statistical errors.
The results are shown in Fig.~\ref{fig:sin}. The minimum $\chi^2$ was
found to be $\chi^2_{min}/N_{d.o.f.}= 2.17/6$. As an additional check
of our global result we have also computed the predicted allowed
values of the weak mixing angle excluding one single experiment at a time
from our global fit. In all the cases the results are similar.

The present analysis of a weak mixing angle generalizes the recent result
$\sin^2\theta_W=0.27\pm0.03$~\cite{Barranco:2005ps} by adding the data from
the LAMPF collaboration~\cite{Allen:1992qe} and from the Krasnoyarsk
reactor~\cite{Vidyakin:1992nf}.
From the global fit shown in Fig.~\ref{fig:sin} we have derived a new
value on the weak mixing angle in the low energy range (below
100~MeV):
\begin{equation}
\label{eq:weak-angle}
\sin^2\theta_W=0.259\pm0.025 
\end{equation}
This value is 1.45 standard deviations larger than the value of the
weak mixing angle obtained from a global fit to electroweak
measurements without neutrino-nucleon scattering data,
$\sin^2\overline{\theta_W}=0.2227\pm0.00037$~\cite{Zeller:2001hh},
that will be used in our numerical calculations of the neutrino charge
radius.

The obtained precision of the weak mixing angle in our analysis is
about 10\% which is close to the expected sensitivities of several
proposals aiming to detect neutrinos in the same low energy
range~\cite{Balantekin:2005md,Scholberg:2005qs,Bueno:2006yq}.

The result~(\ref{eq:weak-angle}) is not competitive with the current
best measurements at low energies obtained from atomic parity
violation~\cite{Bennett:1999pd} and from Moller scattering by SLAC
E158~\cite{Anthony:2005pm} which have a precision better than 1\%. 
However, it is derived from a different channel and it could therefore
give new information about other effects, such as the electron neutrino
effective charge radius.

The gauge-invariant process-independent expression for the Standard Model
neutrino charge radius has been derived in one-loop approximation in
Refs~\cite{Bernabeu:2000hf,Bernabeu:2002pd,Bernabeu:2002nw}
\begin{equation}
\langle r_{\nu_i}^2\rangle_{SM} = \frac{G_F}{4\sqrt{2}\pi^2}\left[
3-2\log\left(
\frac{m_i^2}{m_W^2}
\right)\right]\,,
\end{equation}
where $m_W$ is the $W$-boson mass and  $m_i$ denote the lepton
masses for $i=e, \mu, \tau$.  A numerical evaluation gives the value for
the electron neutrino charge radius squared~\cite{Bernabeu:2000hf}
\begin{equation}
\langle r_{\nu_e}^2\rangle_{SM} = 0.4\times10^{-32}~\mbox{cm}^2\,.
\end{equation}

Besides the standard tree-level amplitude and the $\langle
r_{\nu_e}^2\rangle_{SM}$ contribution, one should keep in mind that
the full neutrino-electron scattering amplitude in one-loop
approximation contains additional
terms~\cite{Bernabeu:2000hf,Bernabeu:2002pd,Bernabeu:2002nw}: the
photon-$Z$ mixing term and the box diagrams involving $W$ and $Z$
bosons.  Therefore, in a single process experiment like the one
considered in our paper, one cannot separate and measure different
contributions. Potentially this can be done by combining data from
several neutrino-electron and neutrino-neutrino scattering processes
as it was discussed in Ref.~\cite{Bernabeu:2002nw}.

We consider therefore an effective approach and, following previous
literature~\cite{Grau:1985cn,Vogel:1989iv,Hagiwara:1994pw}, we denote the
vector coupling as 
\begin{equation}
g_V = 1/2+2\sin^2\theta_W +
(2\sqrt{2}\pi\alpha/3G_F)\langle r_{\nu_e}^2\rangle\,,
\end{equation}
where $\langle r_{\nu_e}^2\rangle$ contains all the contributions discussed
above.  The expression translates into an effective displacement in the value
of the weak mixing angle $\sin^2\theta_W = \sin^2{\overline{\theta_W}} +
\delta $ with the radiative correction $\delta=(\sqrt{2}\pi\alpha/3G_F)\langle
r_{\nu_e}^2\rangle=2.3796\times10^{30}$~cm$^2\times\langle
r_{\nu_e}^2\rangle$~\footnote{A different notation is used in some articles
  ($g_V = {\bar g_V} + \delta$) that implies a definition of $\langle
  r_{\nu_e}^2\rangle$ smaller by a factor two.}.
In the last expression we have taken the fine
structure constant, $\alpha$, and the Fermi coupling constant, $G_F$, as
reported by the Particle Data Group~\cite{Yao:2006px}.
As previously referred, $\sin^2{\overline{\theta_W}}$ is the value of
the weak mixing angle without taking into account the contribution
from $\langle r_{\nu_e}^2\rangle$.
\begin{figure}
\includegraphics*[width=\columnwidth]{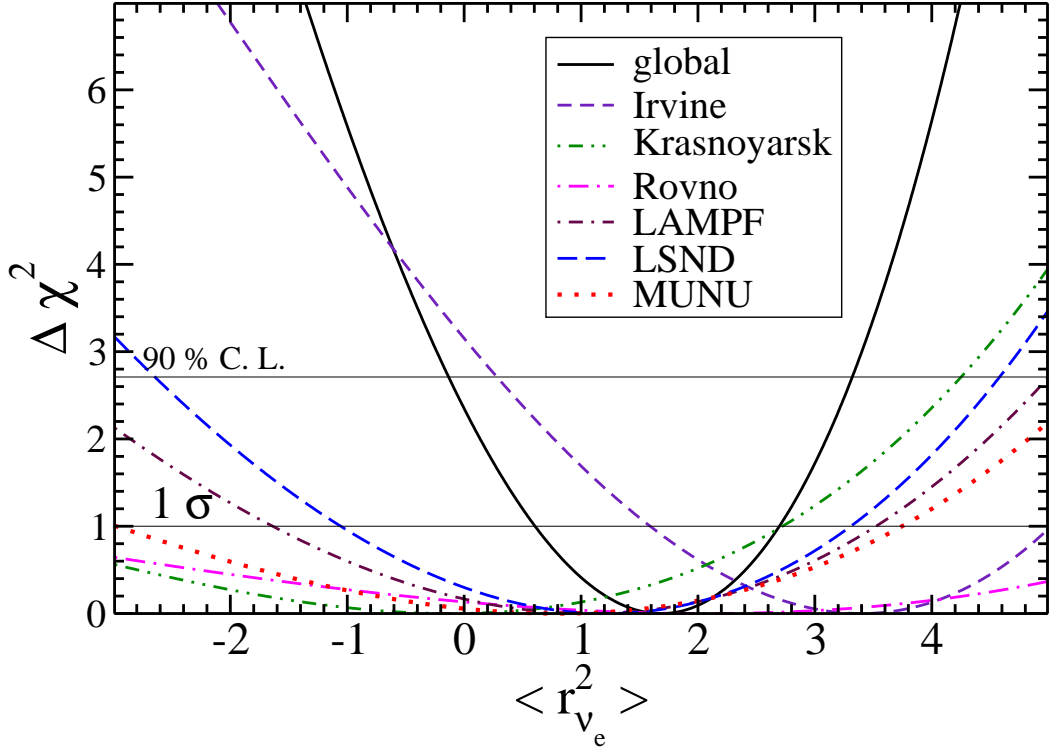}
\caption{Global $\Delta \chi^2$ for electron neutrino charge radius,
  $\langle r_{\nu_e}^2\rangle$ in units of $10^{-32}\mbox{cm}^2$,
  obtained from all discussed $\nu_e e$ and $\bar{\nu}_e e$ scattering
  experiments is plotted. The contribution of each experiment to the
  $\Delta \chi^2$ is also shown.
}\label{fig:necr}
\end{figure}
We have done a $\chi^2$ analysis for every experiment and combined the results
into a global fit as shown in Fig.~\ref{fig:necr}. The obtained allowed
region for the electron neutrino effective charge radius is
\begin{equation}
\label{eq:radius}
 -0.13 \times 10^{-32}\mbox{cm}^2 < \langle r_{\nu_e}^2\rangle < 
3.32 \times 10^{-32}\mbox{cm}^2\quad\mbox{at 90\% C.L.,}
\end{equation}
or $\langle r_{\nu_e}^2\rangle = 1.69_{- 1.09}^{+ 1.01} \times
10^{-32}\mbox{cm}^2$ for one standard deviation.  Although the
precision of present experimental measurements is not enough to
conclude that electron neutrino charge radius is non-zero, better
sensitivity is expected in future proposed experiments and they will
be discussed below.

In Table~\ref{table:exp} we have shown the list of experiments along
with their cross section measurements, the obtained weak mixing angle
and the neutrino charge radius constraint, if available. Our combined
global result for weak mixing angle and electron neutrino charge
radius is also shown for comparison.

\begin{center}
\begin{table}[h]
\begin{tabular}{|p{2.5cm}|p{1.7cm}|p{1cm}|p{3.2cm}|c|c|c|}
\hline
Experiment & Energy & Events & Measurement, $\sigma$ & $\sin^2\theta_W$ & 
$\langle r_{\nu_e}^2\rangle>$ & $\langle r_{\nu_e}^2\rangle<$\\
\hline\hline 
LAMPF~\cite{Allen:1992qe} & 7-60  & $236$ &
$[10.0\pm1.5\pm0.9]
E_{\nu_e}\cdot10^{-45} {\rm cm}^2$  & $0.249\pm0.063$ & $-3.56$ & $5.44$ \\[.1cm]
LSND~\cite{Auerbach:2001wg} & 10-50  & 191 &
$[10.1\pm1.5]\cdot
E_{\nu_e}\cdot10^{-45} {\rm cm}^2$  & $0.248\pm0.051$ & $-2.97$ & $4.14$ \\[.1cm]
 & 1.5-3.0  & 381 &
$[0.86\pm0.25]\cdot \sigma_{V-A}$  & & & \\[-.6cm]
Irvine~\cite{Reines:1976pv}~$\left\{\rule{0cm}{0.7cm}\right.$ & & & &
$\left.\rule{0cm}{0.7cm}\right\}0.29\pm0.05$ & N/A & N/A \\[-.6cm]
 & 3.0-4.5 & 77 &
$[1.7\pm0.44]\cdot \sigma_{V-A}$   & & & \\[.1cm]
Krasnoyarsk\cite{Vidyakin:1992nf} & 3.15-5.175 & N/A &
$[4.5\pm 2.4]\cdot10^{-46} {\rm cm}^2 / {\rm fission}$ & 
$0.22_{-0.8}^{+0.7}$ & $-7.3$ & $7.3$ \\[.1cm]
Rovno~\cite{Derbin:1993wy} & 0.6-2.0 & 41 &
$[1.26\pm 0.62]\cdot10^{-44} {\rm cm}^2 / {\rm fission}$ & N/A & N/A & N/A \\[.1cm]
MUNU~\cite{Daraktchieva:2005kn} & 0.7-2.0  & 68 &
$1.07 \pm 0.34$ events day$^{-1}$  & N/A & N/A & N/A \\[.1cm]
\hline
\hline
Global & & & & $0.259\pm0.025$ & $-0.13$ & $3.32$ \\[.1cm]
\hline
\end{tabular}
\caption{Current experimental data on electron-(anti)neutrino electron
  scattering including measurements of $\sin^2\theta_W$ and limits on $\langle
  r_{\nu_e}^2\rangle$ given by the collaborations. 'N/A' means that a
  collaboration hasn't published the value. Note that MUNU has provided 
  number of events, but not cross-section. Global limits are obtained from
  the  combined analysis of cross-section measurements of all experiments. 
  Neutrino effective charge radius, $\langle r_{\nu_e}^2\rangle$, 
  limits are in units of $10^{-32}$~cm$^2$ for 90\% C.L. range. 
  All energies are in MeV.}\label{table:exp}
\end{table}
\end{center}

\section{Future reactor experiments}

Finally we discuss the perspectives to improve the neutrino charge
radius constraints.  It was recently proposed that future reactor
experiments aiming to measure the neutrino mixing angle $\theta_{13}$
will be able to reach a very good sensitivity to neutrino-electron
scattering cross section~\cite{Conrad:2004gw}. Therefore the weak
mixing angle could be measured with 1\% precision at very low energies
with reactor neutrinos.

We have estimated the sensitivity of such measurements to the electron
neutrino charge radius. In order to get our estimates we have
considered a fixed sensitivity to the weak mixing angle and derived
the corresponding resolution for a neutrino charge radius.  The
results are given in Table.~\ref{table:future}.

\begin{table}[h]
\begin{tabular}{|c|c|c|}
\hline
Experimental &   $1\sigma$ sensitivity  &    $3\sigma$ sensitivity\\
$1\sigma$ resolution  &   to $\langle r_{\nu_e}^2 \rangle$  &   
to $\langle r_{\nu_e}^2 \rangle$  \\
of $\sin^2\theta_W$  &   in $10^{-32}$~cm$^{2}$   &    
in $10^{-32}$~cm$^{2}$\\

\hline
 10\% (present)  &  1.05 & 2.9 \\
 5\%     &       0.47  &   1.4  \\
 3\%     &       0.29  &   0.84 \\ 
 1\%     &       0.10  &   0.28 \\
\hline
\end{tabular}  
\caption{Present (this paper) resolution to weak mixing angle,
  $\sin^2\theta_W$, and the corresponding $1\sigma$ and $3\sigma$ sensitivity
  ranges to $\langle r_{\nu_e}^2 \rangle$ (in units $10^{-32}$~cm$^{2}$) are 
  shown. Expected future sensitivity is given for different values (1\%, 3\% 
  and 5\%) of estimated $\sin^2\theta_W$ resolution.}
\label{table:future}
\end{table}

One can conclude, that from the measurement of the weak mixing angle with 1\%
resolution it is possible to find a strong evidence for electron neutrino
charge radius, which is estimated theoretically to be of the order of
$0.4\times10^{-32}$~cm$^2$~\cite{Bernabeu:2000hf}.

\section{Summary}

We have obtained the weak mixing angle with 10\% precision at energies
below 100~MeV from (anti)neutrino electron scattering off
electrons. To get this result we have combined all available data from
accelerator (LSND and LAMPF) and reactor (Irvine, Rovno, Krasnoyarsk
and MUNU) experiments.

This analysis was also used to set a new limit to the electron
neutrino effective charge radius squared which improves previously
published bounds~\cite{Yao:2006px}.

Future reactor experiments with the estimated precision of
$\delta(\sin^2\theta_W) \sim 1\%$ will be able to find a strong
evidence for theoretically predicted electron neutrino effective
charge radius. 

\section*{Acknowledgments}
  This work has been supported by CONACyT, SNI-Mexico and PAPIIT  project 
  No. IN113206.  TIR was supported
  by the Marie Curie Incoming International Fellowship of the European
  Community and he also acknowledges partial support from the Russian
  foundation for basic research (RFBR) and from the RAS Program ``Solar
  activity''.  TIR thanks Physics Department of CINVESTAV for the hospitality
  during the visit when part of this work was done.

\end{document}